\documentclass[conference]{IEEEtran}

\usepackage{graphicx}
\usepackage[hyphens]{url}
\DeclareUnicodeCharacter{21E4}{}

\ifCLASSINFOpdf
\else
\fi
\begin{document}
%
\title{Holistic Privacy and Usability of a Cryptocurrency Wallet}

\author{\IEEEauthorblockN{Harry Halpin}
\IEEEauthorblockA{Nym Technologies\\
harry@nymtech.net}}


%


\maketitle

\begin{abstract}
  In this study, we overview the problems associated with the usability of cryptocurrency wallets, such as those used by ZCash, for end-users. The concept of ``holistic privacy,'' where information leaks in one part of a system can violate the privacy expectations of different parts of the system, is introduced as a requirement. To test this requirement with real-world software, we did a 60 person task-based evaluation of the usability of a ZCash cryptocurrency wallet by having users install and try to both send and receive anonymized ZCash transactions, as well as install a VPN and Tor. While the initial wallet installation was difficult, we found even a larger amount of difficulty integrating the ZCash wallet into network-level protection like VPNs or Tor, so only a quarter of users could complete a real-world purchase using the wallet.
\end{abstract}


%

\section{Introduction}

The lack of privacy on public blockchains has become one of the most pressing issues for users of cryptocurrency. In response to this concern, a number of privacy-enhanced cryptocurrencies have been developed that store the information ``on-chain'' (i.e. on the public blockchain) in a privacy-respecting manner that attempts to make the sender and receiver anonymous to third-parties that can read the blockchain. Yet the space of these privacy-enhanced cryptocurrencies is absurdly fractured as each of them uses a different cryptographic technique to obtain some level of anonymity for their users, ranging from the zkSNARKs (Zero Knowledge Succinct Arguments of Knowledge) of ZCash~\cite{zcash} to homomorphic encryption of Confidential Transactions~\cite{back2014enabling}, or even the ring signatures in Monero~\cite{noether2015ring}. While these technical debates are of interest, they ignore one real issue that persists regardless of the precise cryptographic primitive used to preserve user privacy: Are these anonymous cryptocurrency wallets even usable by highly-motivated cryptocurrency users? 

While debates rage about the best anonymous cryptocurrency, users themselves are looking to enable this advanced privacy feature via a ``wallet,'' the software used to send and receive cryptocurrency. However, these wallets are still limited in defending user privacy, as they are not integrated with the protection of IP addresses using technologies VPNs or Tor.\footnote{At the time of the study, no ZCash wallet supported Tor integration, came with a VPN, or used a technique like Dandelion to obscure the IP address~\cite{fanti2018dandelion}.} Therefore, it should be expected that users become confused over the precise privacy properties provided to them by ``anonymous'' cryptocurrency wallets.

In order to encourage adoption of tools that match the privacy expectations of users, usability testing is needed on existing real-world cryptocurrency wallets that let users send and receive privacy-enhanced cryptocurrency. However, currently no usability testing has been done on these wallets. Users find themselves confronted with a bewildering choice of wallets and confusing configuration instructions. Even the desiderata for what an anonymous cryptocurrency wallet entails is not agreed upon, leaving users to their own devices when trying to understand the privacy properties of their wallets and transactions. 

After reviewing relevant background literature in Section~\ref{background},  we overview the needs of users in terms of \emph{holistic privacy} in Section~\ref{holistic},  briefly describing  the different privacy-enhanced cryptocurrencies and available wallets  and their problematic relationship to network-level privacy protection. Then we design a series of task-based usability experiments for the ZCash wallet ZEC-wallet Lite in Section~\ref{experiment}.  Section~\ref{results} iterates through the results, with next steps outlined in Section~\ref{conclusions}.

\section{Background}
\label{background}

\subsection{Threat Model and Privacy Properties}

In order to understand the usability concerns of privacy-enhanced cryptocurrency users, there needs to be an understanding of the users' perceptions of their threat model. In general, the threat model is that of a \emph{global passive adversary} that can observe, but not alter, any of the transactions in the network and can observe the entire network. Although this adversary may seem too powerful to be realistic, given that the blockchain records all cryptocurrency transactions into a public ledger, this adversary is realistic insofar as anyone may download and inspect the blockchain. Although the blockchain does not include metadata like the IP address, this information is public in the peer-to-peer broadcast of the chain. As these peer-to-peer networks are fairly small (consisting of ten thousand full nodes for Bitcoin\footnote{\url{https://bitnodes.io/}} but less than a thousand for ZCash\footnote{\url{https://explorer.zcha.in/network}}) so a sybil attack to observe network information such as IP addresses is realistic. Take the case of Chainalysis, a private cryptocurrency forensics company, for example: Chainalysis was revealed to be running over 250 Bitcoin nodes in 2015 in order to observe the IP addresses of Bitcoin users.\footnote{\url{https://www.coindesk.com/chainalysis-ceo-denies-launching-sybil-attack-on-bitcoin-network}}

  Privacy is a vague term capable of having many definitions~\cite{kuhn2019privacy}, but the particular property that is meant to be provided by privacy-enhanced cryptocurrencies such as ZCash and Monero is \emph{unlinkability} to \emph{third parties}, namely that an adversarial observer gains no information from observations of items of interest in terms of the sender and receiver~\cite{pfitzmann2001anonymity}.  For example, if an third-party observer notices that a cryptocurrency transaction on a blockchain is done at the same time as a purchase from a website, an outside observer can infer that the transaction is likely linked to the purchase. These particularly privacy-enhanced cryptocurrencies such as ZCash and Monero can be termed with more precision \emph{anonymous cryptocurrencies} insofar as they in theory can possess \emph{third-party anonymity} as regards global passive adversaries that are monitoring the blockchain in order to discover the sender, receiver, and amount sent as well as \emph{sender anonymity} in terms of who is sending a transaction (as the sender is identified as a pseudonymous address). 

  There has been considerably less work on attempting to understand what threat models and privacy properties actually match the intuitions and expectations of users. While there have been no prior usability studies of anonymous cryptocurrencies, large-scale studies of at-risk activists that show that privacy is a core concern, and that their understanding of privacy - while sometimes confused with the more narrow use of encryption - does indeed map to a global passive adversary capable of observing the metadata of their communication~\cite{halpin:coordination}. This threat model is also mentioned explicitly in the ZCash (Zerocash) design, which notes that ``Zerocash only anonymizes the transaction ledger. Network traffic used to announce transactions, retrieve blocks, and contact merchants still leaks identifying information (e.g , IP addresses). Thus users need some anonymity network to safely use Zerocash''~\cite{zcash}. Therefore, allowing users to enable third-party anonymity in terms of unlinkability between transactions will be the goal of this user study.

\subsection{Usability of VPNs and Tor}

Usability ends up being crucial to the real-world adoption of privacy-enhancing technology for anonymity as ``usability for anonymity systems contributes to their security, because usability affects the possible anonymity set. Conversely, an unusable system attracts few users and thus can't provide much anonymity''\cite{dingledine2005challenges}. Unlike many other types of systems, anonymous systems like Tor or ZCash needs many users, and thus usability, in order to achieve their privacy property of unlinkability between transactions.

One of the fundamental components of internet connections is the Internet Protocol (IP) address of a device, and this information can be used to link user transactions. IP address information can be used to censor information like streaming media by geolocation and even can also be used as an identifier to target advertising and even arrest users.\footnote{\url{https://iranthreats.github.io/resources/webrtc-deanonymization/}} In order to prevent censorship and targeting, many users use VPNs (Virtual Private Networks) where all of their internet traffic is routed through another server (usually in another jurisdiction) so that their identifying IP address is obscured insofar as all their outgoing traffic will have the IP address of the VPN server at its destination. The VPN market is one of the largest in the world, with approximately one of five internet users having used a VPN within the last month.\footnote{\url{https://www.statista.com/statistics/306955/vpn-proxy-server-use-worldwide-by-region/}} Yet VPNs have been heavily understudied, with only a few results from the academic security literature either by noting the poor security of existing VPN providers~\cite{khan2018empirical} or promoting a more secure new VPN protocol~\cite{donenfeld2017wireguard}. From a usability perspective, there have been studies putting forward more usable VPNs (but lacking user studies)~\cite{leap} and the usability risks of VPNs on mobile devices in terms of privacy~\cite{ikram2016analysis}. Perhaps because VPNs have a very weak threat model and are fragmented in terms of user experience, it appears the majority of usability studies have moved to the studying the Tor network~\cite{dingledine2004tor}.

  The Tor network improves on VPN by using encryption with  a peer-to-peer set of relays to collect a circuit of at least three ``hops'' so that the entry to the Tor node, which knows the IP address of the user, does not know the destination~\cite{dingledine2004tor}. However, the network has evolved considerably since early usability studies of Tor, which noted that privacy-enhancing technologies of Tor needed a better conception of their end-user needs in order to be more usable and created the now standard guidelines for privacy-enhanced user testing~\cite{clark2007usability}. Later studies have shown via small scale qualitative interviews that while expert users had a good understanding of Tor, non-expert users believed ``intelligence agencies, such as the Central Intelligence Agency (CIA) and the National Security Agency (NSA), are capable of defeating the protection Tor provides''~\cite{gallagher2017new}. Of course, one should not dismiss these fears by non-expert users, as targeted attacks by the NSA on the Tor browser were revealed by Snowden, and large-scale data analysis of Tor entry and exit nodes can lead to the correlation of traffic patterns and so identification of the users~\cite{ShWa-Timing06}. Other work has focused on Tor's usage for censorship circumvention~\cite{lee2016tor}, the difficulty of using Tor on mobile devices~\cite{assal2014will}, and the negative effect of latency on the usability of Tor~\cite{fabian2010privately}.  

\subsection{Usability of Bitcoin wallets}

Bitcoin is a digital currency where users are identified with their key material, and use cryptographic operations to transfer value -- or even other kinds of data -- into a distributed ledger that records their transactions and simultaneously maintains a consensus to the order and kinds of transactions~\cite{bonneau2015sok}. Addresses are often given as the hash of a public key (with new keys per transaction being generated per transaction with Bitcoin), and private key recovery is done via long mnemonic passphrases.  At first glance, this seems to be a very user-hostile design, as historically in other kinds of systems, such as the encryption of email via Pretty Good Privacy (PGP), there has been a large amount of confusion due to the lack of usability of key material~\cite{johnny}. Yet despite sharing the same fundamental metaphor of a user being identified by a hash of key material with PGP, Bitcoin wallets have obtained a measure of popularity far outstripping PGP, with an estimated 40 million wallets being active.\footnote{\url{https://www.statista.com/statistics/647374/worldwide-blockchain-wallet-users/}} Indeed, although users find using keys to be difficult at first, the usage of keys for financial accounts is a metaphor that Bitcoin users that are typically managed by custodial ``web-based'' wallets where a company like Coinbase holds and manages keys on behalf of users~\cite{kazerani2017determining}. Over time, users are eventually  are comfortable with using keys and wallets that they control~\cite{eskandari2018first}. Previous user studies of relatively small samples of users (for example, 22 users) show that users still find credit cards more usable and are often confused by Bitcoin wallets, even though some features such as the use of QR codes are considered more advanced than credit cards~\cite{alshamsi2019user}.

\section{Holistic Privacy}
\label{holistic}

\subsection{What is Holistic Privacy?}

What has not been studied in previous usability studies is the relationship of privacy features such as anonymity to cryptocurrency. This is due to the inherent social complexity of privacy when combined with the technical complexity of cryptocurrency. There is some useful previous research in cryptocurrencies about user's \emph{perceptions} of privacy features like anonymity in Bitcoin. In particular, studies show that ``32.3\% of our participants think that Bitcoin is per-se anonymous while it is in fact only pseudonymous. 47\% thinks that Bitcoin is not per-se anonymous but can be used anonymously. However, about 80\% think that it is possible to follow their transactions''\cite{krombholz2016other}. While these results seem counter-intuitive, it shows that there is some belief in the privacy features of Bitcoin (even if the details are confused), although users are rightfully concerned that attempts to use Bitcoin anonymously are likely to be defeated by a powerful adversary.

Unlike encryption that -- given an mathematical hardness assumption -- can be proven to keep confidentiality against even powerful enemies,  privacy is not a binary property but a multi-dimensional and socially-embedded spectrum that can be realized in multiple ways, ranging from pseudonymity to anonymity. This especially holds true for third-party anonymity in technical systems, which have multiple layers where information may leak  and so allow an adversary to link items of interest. A technical and philosophical framework that is capable of understanding these various kinds of information leaks it the concept of \emph{levels of abstraction}~\cite{floridi}. When typically describing a complex architecture, the system is divided into discrete levels, where each layer has properties that are inherited from the prior level. This framework is often inherit in technical designs, such as the distinction between the ``application'' and ``network'' layer in the Open Systems Interconnection model (OSI model).\footnote{\url{http://standards.iso.org/ittf/PubliclyAvailableStandards}}

The problem then facing the design of a usable anonymous cryptocurrency wallet is \emph{holistic privacy}: Privacy is not reducible to a single layer of abstraction, and a leak at any layer can eliminate any privacy properties of the system as a whole. This problem is not confined to privacy-enhanced cryptocurrency wallets, but the hard problem of holistic privacy is  endemic in the development of privacy-enhancing tools from web browsers like Tor to secure messaging applications like Signal. For example, Tor cannot interoperate with the popular Google Chrome browser due to the way it retrieves DNS without allowing a proxy like Tor, and Signal uses phone numbers for identities in a way that many activists feel uncomfortable with even though the Signal server itself uses advanced techniques so that it does not need to record the phone number itself~\cite{halpin:coordination}. Holistic privacy is likely always incomplete, as new bugs and layers of abstraction are discovered and more complex assemblages of programs created (in turn, altering their original privacy properties).

\subsection{The Holistic Privacy of Anonymous Cryptocurrency}

Applying this concept of holistic privacy to anonymous cryptocurrency, one needs to divide the usage of cryptocurrency wallets and protocols into distinct levels and inspect each level for privacy leakage. 

There are multiple layers of technology outside of the blockchain itself that must be considered in anonymous cryptocurrency wallet design: In developer parlance, the peer-to-peer networking level is called ``Layer Zero'' while the blockchain level itself is ``Layer One,'' and payment, scaling, and application and off-chain channel levels are called ``Layer Two.'' This makes designing and even using anonymous cryptocurrency difficult, as a leak of information on any layer may de-anonymize a user's transaction on the rest of the layers. There is also some evidence that users are aware of this multi-layer nature of privacy in cryptocurrency transactions, as recent usability studies have shown that ``25\% reported to have used Bitcoin over Tor to preserve their privacy''\cite{krombholz2016other}.  This analysis in terms of levels is important, as even zero-knowledge proofs on the blockchain that can be rigorously proven not to leak any information on chain and obtain the maximum anonymity set per transaction (i.e. the set of all users that ever used the chain) have been shown to be vulnerable to ``side channel'' attacks on other components of the network~\cite{tramer2020remote}. As levels inherit the problems of prior levels, the ``layer zero'' of networking is the place to begin any experiment as levels above inherit the problems of those levels. 

Beneath the blockchain itself that records transactions, the entire Bitcoin blockchain is built on top of a peer-to-peer broadcast layer that sends and receives transactions, similar to classical peer-to-peer systems also studied in the usability literature~\cite{good2003usability}. On the level of networking, Bitcoin broadcasts TCP/IP packets (or UDP packets in some other blockchains) to all other nodes in the network.

This opens Bitcoin to a number of attacks, as these nodes are sent using a simple gossip protocol and are not encrypted, and so may be intercepted by an adversary~\cite{biryukov2019privacy}. Unless advanced techniques are used like Dandelion~\cite{fanti2018dandelion}, these packets will contain the IP address of the sender, and so can be used to de-anonymize the sender regardless of the payload that will contain information stored on the blockchain (i.e. ``on chain'') for a given transaction. Simply using Tor has a host of problems, such as repeated transactions reusing the same IP address\cite{biryukov2015bitcoin}. As this kind of networking information also reveals the approximate time of transaction, and so combined with flaws in the wallet itself has led to attacks on even privacy-enhanced cryptocurrencies such as ZCash that can de-anonymize information that would be otherwise be considered anonymized on-chain~\cite{tramer2020remote}.

There are also layers ``above'' the chain, consisting of applications. For example, popular custodial wallets used by cryptocurrency exchanges like Coinbase often include user-centric information such as the ``Know Your Customer'' information needed for compliance with regulatory frameworks for transferring cryptocurrency and converting cryptocurrency to fiat~\cite{kazerani2017determining}. This includes the vast majority of wallets for privacy-enhanced cryptocurrencies like ZCash that are stored on exchanges such as Gemini.\footnote{\url{https://gemini.com/blog/gemini-is-now-the-worlds-first-licensed-zcash-exchange}} Even if a cryptocurrency itself is anonymous, the methods of accessing it and using it within applications may not be anonymous. This is even true for decentralized applications (dApps), such as the use of the ``Tornado Cash'' mixer\footnote{\url{https://tornado.cash}} on Ethereum, as well as `off-chain' for payment channel solutions meant to scale Bitcoin such as the Lightning Network~\cite{kappos2020empirical}. 

\subsection{The Lack of Privacy-Enhanced Wallets}

There have been many attempts to make user-friendly `anonymous' wallets, but they remained incomplete until recently. Even before the first user-friendly Bitcoin wallets existed, there were concerns with its lack of anonymity. One of the earliest adopters of Bitcoin, Hal Finney, even stated that he was ``looking at ways to add anonymity to Bitcoin.''\footnote{\url{https://www.coindesk.com/hal-finney-bitcoin-words}} In 2014 the DarkWallet project started with the goal of making a user-friendly privacy-enhanced wallet.\footnote{\url{https://www.wired.com/2014/04/dark-wallet/}} Although the project featured techniques that were considered revolutionary at the time, such as ``stealth addresses'' and the use of Confidential Transactions (homomorphic encryption), the wallet was never completed. Research then moved to using new forms of signatures, such as the use of ring signatures of Monero~\cite{noether2015ring} and the use of zero-knowledge proofs by ZCash~\cite{zcash}.

Yet between the launch of ZCash in 2016 and 2020, there was little progress in making usable wallets with a graphic user interface. Only recently, wallets that appear usable for Bitcoin that feature Confidential Transactions, such as the mobile-friendly Blockstream Green wallet, have been released. However, by default this wallet requires two-factor authentication via a phone number (so leaking identity information) and almost no sites currently use Bitcoin with Confidential Transactions (currently only supported on Blockstream by Blockstream's Liquid sidechain~\cite{back2014enabling}). Monero has the most usage of any privacy-enhanced cryptocurrency, however, it's primary wallet is a desktop wallet with a graphical user interface that requires the entire full blockchain to be downloaded, which requires hours of user time and it does not work on mobile phones.\footnote{\url{https://web.getmonero.org/downloads/}}

\begin{figure}[t]
\centering
\includegraphics[width=0.9\columnwidth]{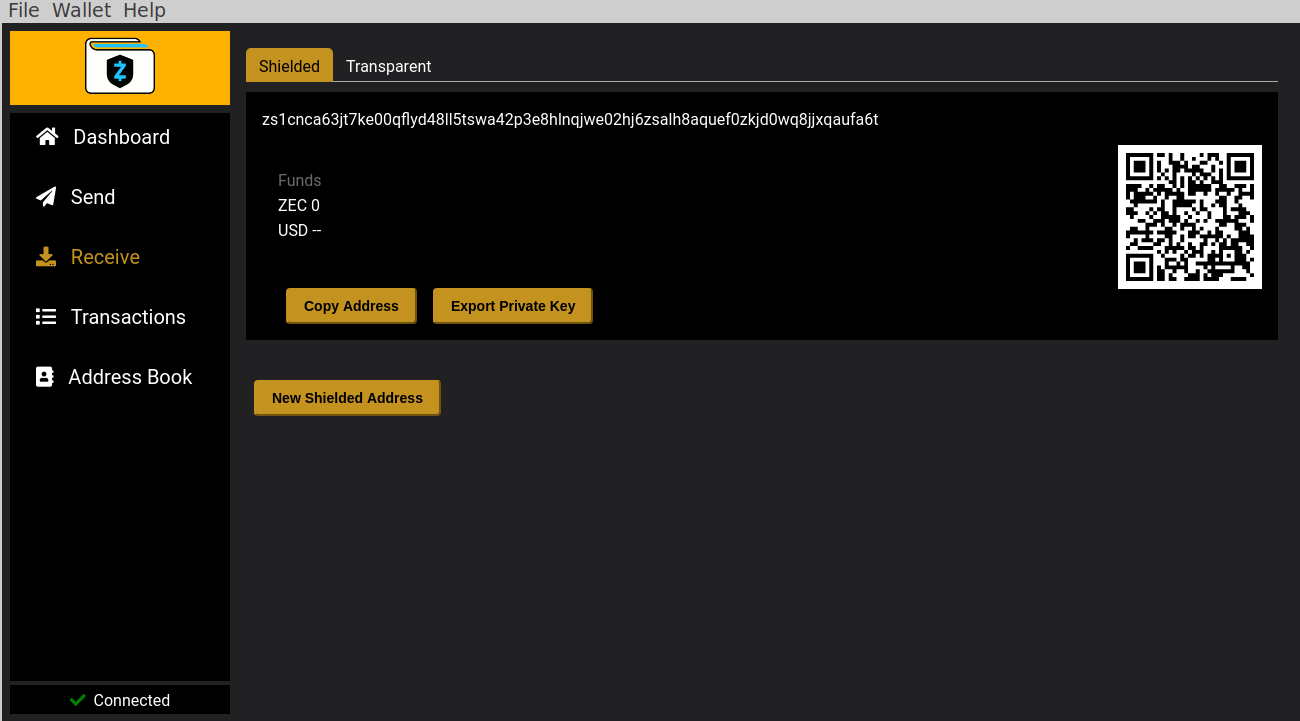}
\vspace{-0.2cm}
\caption{\label{fig:zec} Interface of ZEC Wallet Lite, with a shielded address and equivalent QR code shown.}
\vspace{-0.2cm}
\end{figure}

At the time of the user study (May and June 2020), the only ``light'' wallet (which requires only the hashes to be downloaded, not the full blockchain) that works on mobile with a privacy-enhanced cryptocurrency is ZEC Wallet Lite, which uses ZCash (whose units are called `ZEC').\footnote{\url{https://www.zecwallet.co/}} The interface of ZEC Wallet Lite is shown in Figure~\ref{fig:zec}.  ZEC Wallet Lite allows the use of ``shielded'' ZCash, which uses zero-knowledge proofs to encrypt the transactions on the blockchain with ``shielded'' transactions, allowing transactions that are anonymous to third parties while maintaining auditability. Confusingly, ZCash also supports what are called ``transparent'' transactions, which have no additional privacy and are equivalent to Bitcoin transactions in terms of anonymity. Even more confusingly, ZCash generates two kinds of address types for shielded and transparent addresses that are incompatible (i.e. a shielded address cannot send a transaction to a transparent address and vice-versa, although they are stored on the same ZCash blockchain whose only distinguishing difference from Bitcoin is shielded transactions).

Indeed, the vast majority of usage of ZCash is transparent transactions, with only 6\% of transactions using shielded transactions.\footnote{\url{https://cointelegraph.com/news/zcash-fully-shielded-transactions-jump-70-to-new-record-in-april}} It is unclear why users do not use ZCash shielded transactions, although it may be due to a lack of user-friendly tooling such as ZEC Wallet Lite. More likely, it is due to the fact that vast majority of ZCash transactions are done for speculation via exchanges and the little use of ZCash for anonymity purposes is often easily broken~\cite{kappos2018empirical}. In contrast, Bitcoin and Monero are both very popular in usage at ``dark markets'' where often illegal material is purchased.  ZEC Wallet Lite offers no Tor integration, and due to lack of support of the standard SOCKS5 proxy needed by Tor, it is difficult for even sophisticated users to use ZEC Wallet Lite with Tor.\footnote{\url{https://github.com/adityapk00/zecwallet-lite/issues/26}} 

Based on this analysis, it seems necessary that the desiderata for an anonymous cryptocurrency wallet can be developed. Namely, a user should be able to:
\begin{itemize}
\item Create a new wallet that lets them use an anonymous cryptocurrency
\item Automatically hide networking-layer information
\item Not require any additional personal identifying information to be used by the application
\item Receive and send anonymous cryptocurrency transactions
\end{itemize}

This forms a set of tasks that can be tested for usability. In order to test out the existing state of the art and design new privacy-enhanced cryptocurrencies, a usability study is needed. While ideally these studies would take place in full-blooded social contexts of the users of anonymous cryptocurrencies, or better yet, users who may need anonymous financial transactions but are not yet even aware of the existence of privacy-enhanced cryptocurrencies, more structured studies with groups would provide the first steps as anonymous cryptocurrency software by design should not automatically collect user data without consent, and few users of anonymous cryptocurrency would likely give such consent.

\section{User Study Design}
\label{experiment}

\subsection{Goal}
A user study design was designed in order to understand the expectations and real-world behavior of users when using privacy-enhanced wallets for the common task of paying for items. The principles of the design is based on the primary task of purchasing an item using ZCash privately. The task has some similarities to a ``cognitive walkthrough'' insofar as an expert would illustrate the process and guide users in real-time as much as possible, but different insofar as the expert did not do the task themselves (as this would be less realistic, as non-expert users may have different biases and problems)~\cite{wharton1994cognitive}. In detail, task design was to actually install a new anonymous cryptocurrency wallet, ensure that the user can send and receive transactions, understand how to discover their own IP address and to use a solution to obscure the IP address, and the use this wallet to actually complete a purchase without leaking any non-required personally-identifying information. Realistically, the primary task needed to allow users with a wide variety of experience to participate.

\subsection{Software Choice}

Cryptocurrency wallets either have ``full nodes'' that copy the entire blockchain or ``light'' wallets that copy only the block headers to validate transactions.  We chose to use a ``light wallet'' as otherwise the installation process would be too cumbersome, as takes many hours to install a full node for most mature cryptocurrencies.  As Monero did not offer a ``light'' wallet that supported privacy via Tor or a VPN and user control over keys, so Monero was not used. The same was true of wallets that supported Confidential Transactions. Therefore, despite the inherent confusion in the ZCash design that requires users to distinguish between shielded (anonymous) and transparent (non-anonymous) transactions, ZCash Wallet Lite was chosen.

The website \emph{whatismyipaddress.com} was chosen to help users identify their own IP address during the workshop. ZEC Wallet Lite does not allow integration with Tor in a straightforward manner, as it does support SOCKS5. However, a subtle point should be brought up: Tor does not change the IP address at every transaction. Instead, it develops a bidirectional ``circuit'' (route) through Tor relays that maintains the same proxied IP address (i.e. exit node) for 10 minutes. Furthermore, the generic Tor software does not let new circuits be created per transaction by users easily.\footnote{Although this has been added by the Tails Operating System at \url{https://tails.boum.org}.} In the context of anonymous cryptocurrency transactions, usually a web browser or other application is needed to discover the seller of a product that takes cryptocurrency. Therefore, in order to maintain unlinkability, two IP addresses are needed: 1) One IP address for the web browser and 2) Another IP address for the wallet, as otherwise the IP address of the website could be linked to the IP address of the cryptocurrency transaction. So, a VPN was chosen to hide the IP address of the wallet (i.e. non-HTTP traffic), with \emph{mullvad.net} chosen, while the usage of Tor via the Tor Browser was used for web-surfing to the merchant. The ZCash merchant chosen for the experiment was \url{https://njal.la} that allowed users to purchase, using shielded ZCash, privacy-enhanced servers and domain names as these resources are in high demand by cryptocurrency users. This service was chosen as there are not many services that accept shielded ZCash at the time of the experiment. Finally, we do the experiment only once as repeated purchases would allow the IP addresses to be linked, unless the VPN explicitly restarted a connection with a server with a different IP address or a new Tor circuit with a new IP address was created for the second purchase. 

\subsection{Primary Task}
Before commencing the workshop, the users were debriefed on the goal and structure of the task. Prior, they were given a form to consent to data protection data collection and collect data.  We followed the ethical principles for users as given by Clarke~\cite{clark2007usability}. An expert guided the task, with two other expert guides available to help the participants and verify the step had been completed (as otherwise, a user could leave with the ZCash and, for example, not setup Tor).  Verification via screen-sharing using the guide's custom private videoconferencing (\emph{Jit.si}) server and is done after every main task in the experiment. A large incentive that is twice the amount required for success is given at the end to users who pass the final verification. Although the online setup is more difficult than an in-person experiment,\footnote{However, the online environment was  necessitated by the COVID virus and led to a more diverse user base geographically than would be expected from a typical study that focuses on university students.} the users are given a persistent URL to communicate with the guides over video, so that when their IP address changes they can return to communicate with the guides. The users are given a help forum in Telegram to help continue the experiment at their own pace and a video of the instructions that they can download as well. Users were asked to notify guides if installing the experimental software caused problems to their devices outside of an inability to use the software successfully to complete their task, and no users did so. 

\begin{figure}[t]
\centering
\fbox{\includegraphics[width=0.9\columnwidth]{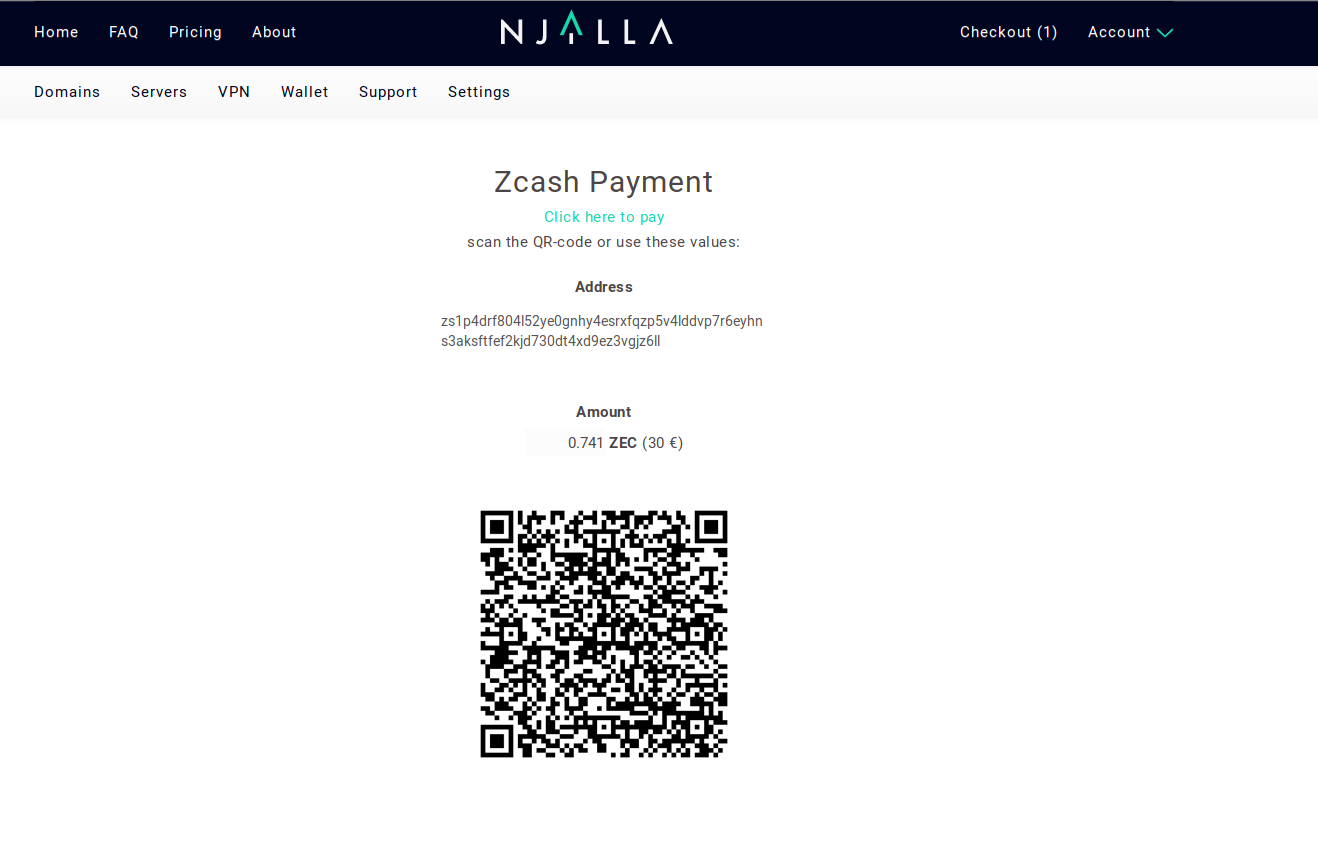}}
\vspace{-0.2cm}
\caption{\label{fig:njalla} Using ZCash for a purchase with a QR code and shielded address. }
\vspace{-0.2cm}
\end{figure}

The walk through had eight main tasks, where each task was illustrated by screenshots in a step-by-step manner with relevant URLs for downloading and using software given at the top of the screenshot:
\begin{enumerate}
\item \textbf{Install the ZCash Wallet Lite}: The user was instructed on the differences between various privacy technologies for cryptocurrencies on a high-level, including the difference between transparent and shielded transactions. Then the user was directed to \url{zecwallet.co} website. A new wallet was created by the user, the user writes down their passphrase and synchronizes the blockchain. The user was asked to share their shielded ZCash address.  
\item \textbf{Receive a ZCash Transaction}: A user was then sent enough ZCash by the experimenter to purchase a server (15 USD in shielded ZEC). The user needs to confirm that they received the transaction to the guide to go to the next step. 
\item \textbf{Discover your own IP address}: The user is redirected to \url{whatismyipaddress.com} and asked of their IP address. 
\item \textbf{Install VPN}:  After an explanation of how IP addresses work, the user is sent to \url{mullvad.net} and walk through the installation process. Access codes are given to allow the VPN to be installed. The user is asked their new IP address by the guides.
\item \textbf{Install the Tor Browser}: The user is redirected to \url{torproject.org} and asked to install the Tor Browser. Then the user is asked for their new web browser IP address by the guides.
\item \textbf{Go to the Merchant website}: The user is redirected to the merchant website of \url{njal.la} and guided through the process of choosing a server to purchase by the guides, although the purchase is not verified at this stage. 
\item \textbf{Register at website using a one-time e-mail address}: The user is given an access code for a one-time e-mail address via \url{riseup.net} and walked through the registration of a new account at \url{njal.la}. The website used by \url{njal.la} is shown in Figure~\ref{fig:njalla}.
\item \textbf{Complete purchase by sending ZCash to merchant}: The user is then instructed to send their shielded ZCash to the website. After the website displays the ZCash address of \url{njal.la}, the user sends the merchant ZCash to purchase a server anonymously. The IP address of the server is then asked by the guides so its purchase can be confirmed. 
\end{enumerate}

At the end of the workshop, users are given approximately twice as much in reward (30 ZEC rather than the 15 needed for the server) in order to incentivize them to purchase the server. Further motivation was given in order to help users use their new domain name and server for tasks such as creating a full-node in a blockchain network.  Any users that eventually succeeded after the formal end of the online workshop were included in the results as ``not succeeding.''

\subsection{Open-Ended Questions}

Users were also given open-ended questions to allow the results to put into more context. The first question was to ascertain their motivation and threat model. The second question went into their prior experience or beliefs about cryptocurrencies, and the final with network-level privacy.  The questions were as follows:

\begin{enumerate}
\item Why do you want to be anonymous? What can you do with anonymous cryptocurrency? What are you afraid of?
\item What anonymous cryptocurrency do you use, or want to use, and why does it make you anonymous?  
\item Do you use a VPN or Tor? Why and what does it do?
\end{enumerate}

The first question was given before the experiment began, the second after task 2, and the last after task 8. Users were allowed to answer questions both in text and voice form. 

\section{Results}
\label{results}




\begin{table*}
\begin{center}
\begin{tabular}{|l|c|c|c|} \hline\hline
\textbf{Goal of Task} &             \textbf{Success} & \textbf{Task No.} \\ \hline\hline
\emph{Received ZCash} &            27 (45\%)       &  1-2    \\ \hline
\emph{Installed VPN and Tor} &   32 (53\%)   &   3-4         \\ \hline
\emph{Purchased using ZCash} &   17 (28\%)   &   6-8         \\ \hline\hline
\end{tabular}
\end{center}
\caption{Successful Completion Per Task}
\label{tab:results}
\end{table*}

\subsection{Participant Selection and Background}

An initial run through of the user study was done in May 2020 in the form of a hands-on workshop with over 100 attendees at the `Consensus Distributed' conference, but the results were not used in this version as data was not collected and the tasks refined. Later in May and June 2020, three sessions of user experiments were run. Users were recruited via spreading a link to a form via Twitter and Telegram via cryptocurrency-specific discussion channels. When clicking the link, the users asked to fill out an online survey and were then contacted via email to participate in one of three user study sessions.  It should be noted these users are highly motivated and so it can be argued highly motivated users who are aware of privacy are not representative of ordinary users. However, we would argue that motivated users actually are a \emph{more representative sample} and better in terms of \emph{understanding the behavior of early adopters of a system} than some mythical `ordinary user,' as this idea is operationalized often in usable security and human-computer interaction studies with a small (perhaps randomly, but often self-selected) group of students from a university, as has been extensively argued earlier~\cite{halpin:coordination}.

Over all sessions, 60 out of 89 users completed the entire survey and consented to data collection while respecting the General Data Protection Directive (note once 60 users were reached, no more users were added for a session). Each session was ran with the online video teleconferencing \emph{Jit.si} software with the experimenters and users sharing a single teleconference room and chat channel. There were three experimenters -- a primary researcher and two assistants -- overseeing the experiment, one giving slides describing each task with graphics, one taking notes, and then another working with the primary researcher in the forum to monitor user progress in the tasks.

In total, 60 users were recruited for the user study. The average age was 34, with users ranging from the ages of 19 to 70 ($\sigma$=10.22), with only 2 users opting-out of revealing their age. 76\% of the users were male, with 17\% asking not to reveal their gender, and 7\% were women. 30\% of users used Mac OS, 28\% of users used Windows, 25\% of the users used Linux, 12\% used Android, and 5\% used iPhone for the experiment (note that the software was chosen as it was multi-platform). 27\% of the users were from Europe, 12\% from the USA, 17\% from Africa, and 8\% from Asia, with 36\% not revealing their location.  Users were asked to give an average of their familiarity with cryptocurrency, with scores ranging from one (not familiar) to six (very familiar). The average response was 4.4 ($\sigma$=1.3), indicating a fairly advanced knowledge of cryptocurrency. 20\% expressed that ZCash was the anonymous cryptocurrency they were most interested in, with 23\% expressing more interest in Monero (the most popular) and  23\% in Bitcoin Confidential Transactions (the rest were a mix of Ethereum and Bitcoin mixers). Users were also asked their familiarity with VPNs and Tor, with scores ranging from one (not familiar) to six (very familiar). Again, the average score was 3.92 ($\sigma$=1.4), which again shows a fair amount of confidence, although slightly less than with cryptocurrency. Approximately 23\% of users had the most experience and interest in Tor, and more users (25\%) had experience and interest in VPNs. However, the majority (48\%) was interested in more advanced cryptocurrency-specific techniques for network-level protection, with the rest interesting in techniques like Shadowsocks to avoid the Great Firewall of China. The user selection was quite global, with a large representation from Germany but also, surprisingly, from Ghana and Nigeria. However, upon inspection, it appeared the link was shared in some African cryptocurrency groups during a time in which interest in cryptocurrency was growing dramatically. The typical profession was student, although there were also software developers, lawyers, and even a DJ that participating. In general, the primary note of interest was that in terms of disclosure users considered their geolocation to be their most sensitive attribute, followed by their gender, but most users were comfortable revealing an age and email address to participate in the experiment. 

\subsection{Task Success}

The results are summarized in Table~\ref{tab:results}. During the one hour user study session, 27 users managed to install the ZEC Wallet successfully and receive ZCash (45\%) and 32 managed to complete the installation of the VPN (53\%). The reason for the high failure rate on the usage of ZEC Wallet is that some users could not get ZEC Wallet working on their operating system, but managed to get the VPN working (and so more users installed the VPN that the wallet). Then 17 (28\%) users managed to complete the entire exercise. Overall, the results are surprising insofar as the users seemed to overestimate their expertise in actually getting software working, which is likely an effect of being recruited from professional Telegram groups targeting cryptocurrency. However, the group was also self-selected and highly motivated, which provides for the fairly good results for a lengthy walk through process over progressively different tasks. 

Measurement of ZCash wallet installation was taken via the receiving of ZCash from the experimenters. The primary reason (25\%) for the failure to install the ZCash wallet seemed to be difficulty getting it to work on some platforms, in particular iOS and users that did not want to use Google Play. This was caused by lack of ``trust in Google'' and in particular fear over ``Google Services'' that users were worried were ``snooping'' on them, and so alternative installation setups were preferred but difficult to get working. A user reported that it ``takes too long to synchronize the blockchain'' even though a light wallet was used, but this was likely a side effect of slow internet connections by some of the users. Among a minority of users (20\%), there was considerable confusion over the location of the ZCash ``shielded'' address needed to receive the funding, despite the wallet being easy to install and prominently displaying the ZCash wallet. This seemed to be in general the fault of confusion over two different addresses types. Users asked of shielded address whether ``its like a separate thing entirely, forgot how it works though'' and one thought (incorrectly) that ``on the exchange you always get two addresses.'' Also, for six users (10\%), the speed of the initial synchronization (or doing it in the background) caused the user to give up, as the blockchain failed to synchronize their devices. This was likely due to a very slow internet connection. Although cryptocurrency often uses the rhetoric of financial freedom, many users in places like Ghana do not have high speed internet or only have it on their mobile phone.

Measurement of the success of VPN installation was taken to be sending of the new ``hidden'' IP address for both the VPN and Tor. The VPN installation was also involved, and featured a smooth installation process on iOS and via native Debian packages, which resulted in an easier installation for some users. Also, the requirement for the VPN and Tor required simply turning it on and reporting an IP address with no waiting for a synchronization of the blockchain or confirming a shielded ZCash transaction. The task was also slightly easier, as if the installation process was completed then it was straightforward to retrieve the IP address.  However, there were considerable failures in installation for the VPN (33\%) was also due to confusion over the usage of the code to create the free VPN account, with only 7\% failing to install the application at all (again, the culprit seemed to be slow download times). Most users understood that ``an IP address tells everyone who I am'' and so ``we need to get a new IP everytime we connect'' and so ``it made sense that there was different numbers.'' Fewer users experienced issues with installing Tor after they installed the VPN (13\%), although this may be biased by some users dropping out while installing the VPN. Tor users who failed to install Tor did so primarily it seemed due to some restrictions on their computer or trouble with the large amount of binaries required by Tor. One user asked ``Why does Tor take so long to download?'' 

The final purchase of the anonymous server was measured via the ability to connect to the server via the user giving the experimenters the IP address. At this stage, some users (7\%) were lost in the confusing process of either finding or cut and pasting the shielded ZCash wallet of the address to their browser as the merchant had a complex website that required filling up a ``user account'' with ZCash before purchasing (which users found counter-intuitive). The site also offered multiple types of addresses and a few users had trouble finding the correct address (although no users explicitly complained over sending funds to the wrong address to our knowledge). Further feedback included device issues, such as ``I am more used to simple QR codes'' and ``had issues scanning QR code on my browser from one device.'' These issues of poor website design are beyond the scope of wallet testing. However, users (5\%) were also ``not sure'' if their VPN and Tor were on, and one user seemed to check compulsively multiple times and so did not send the funds. Some users transferred funds but could not tell if their server was on, and had difficulty finding the IP address. This understandably upset the users. Impatience over verification time was common even among users who succeeded (10\%), with four users complained that the reception of their shielded ZCash took too long, and two users sent ZCash multiple times to the merchant. One user sent a small test transaction first, before sending the entire transaction. As the transactions were to a ``user account'' on \emph{njal.la}, the funds were not lost to the users.

\subsection{Open-Ended Questions}

All of the open-ended questions went through a qualitative thematic analysis by two people, the author (whom had years of experience in user studies) and an assitant with a background in cryptocurrency that took the notes for the experiment. Each question and some of its answers are dealt with below. Not all users were very expressive, and most preferred text chat to video.

\textbf{What anonymous cryptocurrency do you use, or want to use, and why does it make you anonymous?} In terms of the first question, users generally had a good understanding that Bitcoin was not anonymous, stating that it ``is  pseudoanonymous - if your identity is anyhow linked to your address you are exposed'' and ``transactions are traceable, it is twitter for your money.''  Some users got confused over the difference between various anonymous cryptocurrencies, as one user preferred Monero but thought it was the same as Bitcoin, saying they preferred ``Monero, because it is more mature than ZCash and you can do anonymous transactions in BTC.'' Many users expressed a desire to more easily transfer funds from their non-anonymous cryptocurrencies such as Bitcoin and Ethereum to solutions like ZCash or Monero easily without an exchange. However, some users felt ``safer'' using anonymous cryptocurrencies, and many expressed concerns over surveillance, such  that ``governments are after Bitcoin'' but were worried that they could be tracked using anonymous cryptocurrency as its very usage ``would put me on a list.'' Many users noted that the majority of their funds were in Bitcoin and Ethereum, and would have preferred to use Bitcoin via a mixer that kept the results in Bitcoin.

\textbf{Do you use a VPN or Tor? Why and what does it do?} Users often claimed to be using a VPN or Tor, but usually ``infrequently'' with four users using Tor for all web browsing and eleven users using a VPN for all web browsing. Only one user claimed that they had thought of the need for Tor integration with anonymous cryptocurrency.  Most users could say that there was ``multiple hops'' and had an idea that ``Tor used a p2p net'' and VPNs were centralized. Most users expressed distrust with the VPN, saying that they have ``trust issues''  and simply stating that ``I don't trust VPN providers.''  One user expressed distrust in Tor, as ``doesn't work in China and hasn't in a longtime.'' One user thought there was ``something wrong with Tor'' but no user could articulate the problem and the majority of users felt more comfortable with Tor than a VPN. The most trenchant critique of Tor was that one user said he preferred running a ``VPN himself'' as ``it would be harder for the NSA to track down, as they know all the Tor machines.'' However, many user viewed Tor and VPNs as simple tools to make their transactions anonymous where ``Tor was better than a VPN'' but could not explain why they wanted to install a VPN and Tor in order to not having their cryptocurrency transactions have the same IP address as their web browsing. One user said ``I thought this was automatic, that you got a new address with every page'' and that ``Tor should just be easy with ZCash.'' In general, users did not seem to understand how TCP/IP circuits worked and that new IP addresses could not be easily made for a new transaction automatically. 

\textbf{Why do you want to be anonymous? What can you do with anonymous cryptocurrency?} As for the last question, multiple users expressed they wanted to be anonymous to be ``safe.'' Users were excited that they could purchase items with ZCash, and no user in the experiment had actually used ZCash before the experiment with a real-world purchase. Although, a few users failed to manage to discover how to send transactions correctly due to confusion over the wallet interface and difficulty in ``cut and pasting'' the address or using the QR code given by the website, users expressed the strongest desire to purchase domain names and servers anonymously in order to launch ``new projects that can't be linked to my credit card.'' There was heavy interest in running miners and validators for other chains. One user was concerned ``if I made a lot of money, then everyone would know the IP and hack the box'' and it seemed that users were in general interested in buying computational infrastructure in order to be resistant to unknown future attacks, although more by hackers than nation-state adversaries. The users that did express concerns with nation-states did so more due to censorship and taxes rather than human rights, although a few users expressed general concern over the NSA. One user complained about their telephone company ``owning them'' and  another user said ``The NSA has already pawned me'' and ``all of this [surveillance] is fascist.'' One user perhaps summarized the general mood by stating that ``I have the right to move my money'' but ``I hope my systems ain't broken.''   

\section{Conclusion and Future Work}
\label{conclusions}
The goal of designing privacy-enhanced software should be easy-to-use software that preserves the holistic privacy of users. As our study showed, it is difficult for even highly motivated and technically adept users to use anonymous cryptocurrency wallets at the present moment in a way that preserves their holistic privacy due to the fundamental disconnection of network level and on-chain anonymity.  The issues around network protection require attention, as it is difficult to do correctly today despite the fact that a sizable portion of users is genuinely concerned by these issues and willing to engage in not only the installation, but purchase, of these kinds of tools like VPNs. The complex instructions required to install multiple interconnected pieces of software in order to preserve the holistic privacy requires considerable difficulty and cause a large portion of users to fail a task, even when online guidance is given by a tutor with very precise instructions. Although some failure was caused by confusion in interaction with the merchant, the majority is caused by the wallet and its integration with Tor and VPNs.

The obvious solution is to integrate network-level protection with cryptocurrency wallet that makes all transactions anonymous by default and changes the IP address with every transaction.  Then the user study could be repeated as a control experiment with a fully-integrated wallet that features network-level protection and shielded addresses as the only option. This should be fairly simple to do, yet it has not been done by ZEC Wallet or other wallets. In the original ZCash whitepaper, its actual privacy guarantees are explicitly noted to require that ZCash would be integrated with Tor or even more powerful anonymous communication systems such as mixnets~\cite{zcash}, as the original ZCash paper states:``The most obvious way to do this is via Tor. Given that Zerocash transactions are not low latency themselves, Mixnets (e.g.,Mixminion) are also a viable way to add anonymity (and one that, unlike Tor, is not as vulnerable to traffic analysis)''~\cite{zcash}. More than half a decade since the publication of the original work and four years since the deployment of the ZCash codebase, there is no network-level privacy layer that is integrated into a user-friendly wallet. Even if it would be difficult to allow integration with mixnets today on a per transaction basis, integration with a mix-net like Nym\footnote{\url{https://nymtech.net}} or a onion-routing system Tor should be bundled with the wallet.  Lastly, the wallet itself needs to not only provide on-chain anonymity by default for all transactions, but feature a simplified interface that requires less dependencies on third party configuration options for VPNs and Tor. Randomized connections by ``light'' wallets to full nodes, as done by the Solana network, may be one easy solution that does not require centralized trust in a wallet-selected default for a ZCash full node.\footnote{\url{https://solana.com/}} 

  There are many details that point to concrete usability improvements outside of network integration. Although anonymous cryptocurrency worked for a large percentage of users, there is general confusion over both the distinction between transparent and shielded addresses and how to interface ZCash with adequate solutions, including ones like a VPN, for defending network-level privacy. The use of transparent addresses seems to have been a design flaw in the ZCash design, primarily existing to enable speculation rather than privacy-enhanced usage of ZCash (which is the only market distinction it has in comparison to Bitcoin), so it would make sense that ZCash developers work to eliminate transparent addresses. Speed seems relatively acceptable once the wallet has been synchronized, but not enough for users with very unstable or bad internet connections. Also, the users most concerned with privacy often have issues that prevent them from easily installing the software. Without more usable anonymous cryptocurrency wallets to increase the base of users, the real-world privacy of ZCash is likely to be less than Monero given the relatively low current adoption of shielded ZCash, although the use of shielded ZCash may grow in the future. Users were relatively confused over the technical differences between anonymous cryptocurrencies, and the main desire was to use Bitcoin and Ethereum anonymously rather than a new cryptocurrency.

  Another possible experiment would be to compare with ``layer two'' solutions, but the privacy properties of the Lightning Network (as deployed in Blockstream Green wallet for example)  is also poor. The exception is the Bolt layer two solution which is still not deployed in a wallet but under active development~\cite{bolt}. Ethereum wallets do not have network privacy or SOCKS5 proxies, but primarily function via the often bug-prone attempts to use Metamask via the Tor browser, although the implementation of zero-knowledge proofs on Ethereum or zero-knowledge ``rollup'' approaches may eventually allow anonymous Ethereum wallets.

  The study itself is limited and should be expanded in future work. The success of users may be correlated to their confidence like ours where users are self-selected. The study is limited to ZCash and a particular ZCash wallet, as at the time of the study it was the only light wallet that offered on-chain privacy. Future work could also compare this wallet to other new ZCash wallets, such as Nighthawk.\footnote{\url{https://github.com/nighthawk-apps/nighthawk-wallet-android}}. The same study could also be repeated focusing only on the usability of ZCash wallets without attempting to integrate network-level privacy, or focusing on another privacy-enhanced cryptocurrency such as Monero. The experiment could be repeated in terms of network-level privacy with other cryptocurrencies as well.  Bitcoin wallets like Electrum support network-level privacy via a SOCKS5 proxy that could integrate with Tor or alternatives like I2P and the Nym mixnet, as well as Confidential Transactions support on Blockstream Green.\footnote{\url{https://blockstream.com/green/}} However, Bitcoin still offers very poor on-chain privacy~\cite{biryukov2015bitcoin}, and so a comparison with Bitcoin may make sense as a sort of control group, although the wallet interfaces and primary task would be so different that it would be hard to directly compare those results  to the users trying to use ZCash. The same would hold of a comparison to Bitcoin ``mixers'' based on wallets like Wasabi Wallet.\footnote{\url{https://wasabiwallet.io}} The proper control experiment would be a ZEC Wallet Lite that has network-level privacy built in. One proposed remedy is the ``Stolon'' proposal of the ZCash Foundation. This proposal uses Tor circuits, reset at each transaction, to implement a Dandelion system over broadcast circuits, but it is unimplemented and likely would suffer from side-channel attacks as it would need to switch circuits every transaction.\footnote{\url{https://github.com/ZcashFoundation/zebra/blob/stolon/book/src/dev/rfcs/0006-stolon.md}} A superior method would be to natively integrate a mixnet with a large amount of users, as it would provide per-message privacy unlike circuit-based solutions like Tor and VPNs. Otherwise, a user sending multiple Bitcoin transactions in the same time period (10 minutes) easily de-anonymizes themselves~\cite{biryukov2015bitcoin}. The Nym mix network would not force users to understand the details of privacy of the message layer as it would automatically mix every message independently.\footnote{https://nymtech.net}

    Regardless, future studies should broaden the number and diversity of users and the kinds of wallets. Users also seemed open to using new technologies that would make the process simpler and easier to cognitively understand, including mixnets. Some of the problems, such as reminding users to use anonymous browsing in concert with their wallet in order to prevent timing attacks, are quite subtle and difficult for users to grasp. Still, more work is needed in order to address these issues in a systematic fashion in order to design the next-generation of anonymous wallets to achieve holistic privacy for their users.


\section*{Acknowledgment}

The author would like to thank Alexis Roussel and Theo Goodman for helping conduct the user study and reviewing user data.  Most importantly, the author would not have been able to complete this experiment without the participation of the Nym community and their enthusiasm for privacy.



%

\bibliographystyle{IEEEtrans}
\bibliography{main}

\end{document}